# A machine learning–based classification approach for phase diagram prediction

Guillaume Deffrennes[a,*], Kei Terayama[b], Taichi Abe[c,d], and Ryo Tamura[a,d,e,f,*]

[a] International Center for Materials Nanoarchitectonics (WPI-MANA), National Institute for Materials Science, 1-1 Namiki, Tsukuba, Ibaraki 305-0044, Japan

[b] Graduate School of Medical Life Science, Yokohama City University, 1-7-29, Suehiro-cho, Tsurumi-ku, Kanagawa 230-0045, Japan

[c] Research Center for Structural Materials, National Institute for Materials Science, 1-2-1 Sengen, Tsukuba, Ibaraki 305-0047, Japan

[d] Research and Services Division of Materials Data and Integrated System, National Institute for Materials Science, 1-1 Namiki, Tsukuba, Ibaraki 305-0044, Japan

[e] RIKEN Center for Advanced Intelligence Project, 1-4-1 Nihonbashi, Chuo-ku, Tokyo 103-0027, Japan

[f] Graduate School of Frontier Sciences, The University of Tokyo, 5-1-5 Kashiwa-no-ha, Kashiwa, Chiba 277-8561, Japan

∗ Corresponding authors at: National Institute for Materials Science, 1-1 Namiki, Tsukuba, Ibaraki 305-0044, Japan. E-mail addresses: DEFFRENNES.Guillaume@nims.go.jp and TAMURA.Ryo@nims.go.jp

## Keywords

# Abstract

Knowledge of phase diagrams is essential for material design as it helps in understanding microstructure evolution during processing. The determination of phase diagrams is thus one of the central tasks in materials science. When exploring new materials for which the phase diagram is unknown, experimentalists often try to determine the key experiments that should be performed by referencing known phase diagrams of similar systems. To enhance this practical strategy, we attempted to estimate unknown phase diagrams based on known phase diagrams using a machine learning–based classification approach. As a proof of concept, we focused on predicting the number of coexisting phases across the 800 K isothermal section of each of the 10 ternaries of the Al-Cu-Mg-Si-Zn system from the other 9 sections. To increase the prediction accuracy, we introduced new descriptors generated from the thermodynamic properties of the elements and CALPHAD extrapolations from lower-order systems. Using the random forest method, the presence of single-, two-, and three-phase domains was predicted with an average accuracy of 84% across all 10 considered sections with a standard deviation of 11%. The proposed approach represents a promising tool for assisting the investigator in developing new materials and determining phase equilibria efficiently.

# 1. Introduction

Phase diagrams are widely used as roadmaps to support the development of materials, from their synthesis to their applications. Detailed knowledge of phase equilibria and the underlying thermodynamics enables the establishment of the link between processing and microstructure, which is a crucial step in modern material design [1–4]. Thus, the development of new strategies based on machine learning (ML) to promote more efficient determination of phase diagrams is of significant interest. Recently, two main approaches have been found to be particularly promising.

On the one hand, the combination of high-throughput experimentation (HTE) and ML has attracted attention. HTE techniques usually involve the synthesis of thin-film libraries using combinational sputtering [5–7] or diffusion multiples [8–10]. ML approaches then enable the rapid analysis of large volumes of data from HTE, further automating the phase diagram determination [11,12]. On the other hand, ML predictions can help in reducing the number of experiments required to explore unknown material systems. There have already been several successful studies in which new compounds were discovered based on these predictions [13]. In addition, an uncertainty sampling–based active learning strategy for recommending the most informative experiments to be performed for efficient determination of the phase diagrams has been proposed [14–16] and demonstrated in practice [17].

In this study, we attempt to estimate unknown phase diagrams based on known phase diagrams using a machine learning–based classification approach. Compared with the prediction of a given material property, the classification problem is complicated because a variety of phases may be stable depending on the constituent elements. Two ingenuities were introduced to succeed in this task. (i) Compositional descriptors suitable for phase diagram prediction are proposed. These are generated from the thermodynamic properties of the elements and CALPHAD (CALculation of PHAse Diagrams) extrapolations. (ii) The categories are narrowed by focusing on the number of coexisting phases instead of the phases. Although this is a simplified representation of phase diagrams, it still contains essential information on phase equilibria, and the predictions can thus be relevant for planning experiments. An overview of the workflow of this study is presented in Fig. 1.

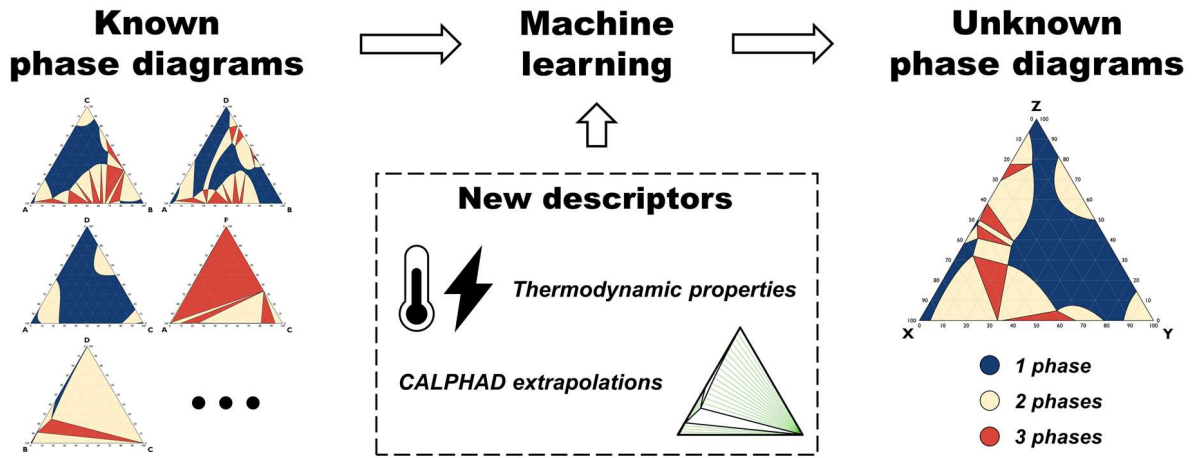

Fig. 1 – Schematic of the phase diagram prediction workflow

As a demonstration, we focus on the constitutive ternary phase diagrams of the Al-Cu-Mg-Si-Zn system. This practical case is selected because reliable thermodynamic descriptions are available for each of these 10 ternaries, which are of significant industrial importance for light alloys. More specifically, the prediction of their 800 K isothermal section is examined. At this temperature, the descriptions are generally supported by many measurements, and the liquid phase begins to stabilize in the majority of systems. To predict ternary sections, the random forest [18] and support vector machine (SVM) [19] classifiers are adopted, which are implemented using Scikit-learn [20].

## 2. Methods

### 2.1. Target ternary phase diagrams

The 10 ternary subsystems of the Al-Cu-Mg-Si-Zn quinary are set as the target for the present investigation. For each phase diagram, the phase equilibria data are retrieved across the entire 800 K isothermal section with steps of 2 at.% using the high-throughput calculation function of the Pandat software [21]. The thermodynamic descriptions adopted in the calculations are as follows: Al-Cu-Mg [22], Al-Cu-Si [23], Al-Cu-Zn [24], Al-Si-Zn [25], Al-Mg-Si [26], Al-Mg-Zn [27], Cu-Mg-Si [28], Cu-Mg-Zn [29], Cu-Si-Zn [30], and Mg-Si-Zn [31]. As a result, 1326 datapoints per system are obtained. For each point, the number of coexisting phases (one, two, or three) obtained by the CALPHAD calculations is used as the label corresponding to the category in the classification problem. In total, the number of single-, two-, and three-phase domains in the dataset is 2263, 6063, and 4934, respectively. The distribution of the classes for each of the 10 isothermal sections is presented in Supplementary Note A (Table S.1).

### 2.2. Compositional descriptors

To predict the phase diagrams with a high accuracy, we use three types of descriptors called "Magpie," "Thermo," and "CALPHAD." These are generated based on the composition of each element. For the "CALPHAD" descriptors, CALPHAD assessments of lower-order systems are necessary, while the properties of the pure elements are used for "Magpie" and "Thermo."

#### 2.2.1. "Magpie" descriptors

For each datapoint investigated, five features are first obtained from the atomic concentration of the elements (at.% Al, Cu, Mg, Si, and Zn). In addition, the mean and absolute deviation of the properties referred to as "Magpie" in Table 1 are calculated from their values for the pure elements and the composition using the Magpie library [32]. This 37-dimensional descriptor set is referred to as the "Magpie" descriptor set. It is noted that this type of descriptor is widely used to predict material properties [13,32].

### 2.2.2. "Thermo" descriptors

The "Magpie" descriptors do not include any thermodynamic properties. However, such information can be expected to be useful for phase diagram prediction. Thus, in the second descriptor set, 10 new related features are introduced. The five thermodynamic properties referred to as "Thermo" in Table 1 are prepared for the pure elements using the SGTE database [33] in its 5.0 version. Their mean and absolute deviation are then calculated at each composition point as for the "Magpie" case. We refer to these 10 descriptors as the "Thermo" descriptor set. A Python script, DesPD (Descriptors for Phase Diagrams), is available at https://github.com/GuillaumeDeffrennes/DesPD to easily obtain these descriptors at any temperature and composition.

### 2.2.3. "CALPHAD" descriptors

At constant temperature, pressure, and composition, thermodynamic equilibrium is characterized by a minimum of the Gibbs energy. Therefore, once this thermodynamic potential is known for all the phases of a system, its phase diagram can be accessed. Within the CALPHAD framework, the Gibbs energy of the phases is modeled based on a building-block approach. For instance, the Gibbs energy of the quinary Al-Cu-Mg-Si-Zn liquid is obtained from the description of the pure elements in the phase, their two-by-two interactions, and their relatively small ternary interactions. This implies that the thermodynamic description of a multicomponent system can be approximated using its subsystems. Herein, we propose the use of the information extrapolated from lower-order systems, i.e., from the binaries in the present case, as descriptors. Specifically, we introduce three features: (i) the number of coexisting phases extrapolated from binary descriptions, (ii) the Gibbs energy extrapolated from binary descriptions, and (iii) the excess Gibbs energy in the liquid phase extrapolated from binary descriptions. These are referred to as the "CALPHAD" descriptor set (Table 1). To obtain this type of descriptor, CALPHAD assessments of lower-order systems (here, the binaries) are necessary. Furthermore, the first two features, i.e., "Extrapolated n_phase" and "Extrapolated Gibbs," have to be calculated using a thermodynamic calculation software, such as those listed in [34]. This is not a requirement for "Extrapolated Ex G Liq" as this descriptor can be obtained via a direct calculation from the interaction parameters (Appendix).

For our demonstration target, the thermodynamic description of the constitutive binaries of the Al-Cu-Mg-Si-Zn system are gathered into a database: Al-Cu [24], Al-Mg [35], Al-Si [36] (based on [37]), Al-Zn [38], Cu-Mg [36] (based on [39]), Cu-Si [23], Cu-Zn [40], Mg-Si [41], Mg-Zn [27], and Si-Zn [25]. On this basis, for each of the 10 ternaries considered in this study, point calculations are performed every 2 at.% across the 800 K isothermal section that is obtained by extrapolation from the binaries. It is emphasized that, when a ternary phase diagram is extrapolated from its binary subsystems, any ternary compounds or solutions of a third element in a binary phase will be missing (these two cases will be simply referred to as ternary phases hereafter). However, a reasonable estimate of the Gibbs energy of the phases that already exist in the unaries (e.g., the liquid phase) can be obtained.

A Python script is available at https://github.com/GuillaumeDeffrennes/DesPD to easily obtain the excess thermodynamic properties in the liquid phase as extrapolated from the binaries.

Table 1 – Compositional descriptors for the phase diagram prediction

| Property | Abbreviation | Statistics | Reference |
|---|---|---|---|
| Atomic percent of Al | Al | Absolute value | Composition (included in the "Magpie" set) |
| Atomic percent of Cu | Cu | | |
| Atomic percent of Mg | Mg | | |
| Atomic percent of Si | Si | | |
| Atomic percent of Zn | Zn | | |
| Atomic number | AtomicNum | Mean (mean) and average deviation (avg_dev) of the property among elements in composition | "Magpie" |
| Mendeleev number | Mendeleev | | |
| Periodic table row | Row | | |
| Periodic table column | Column | | |
| Atomic weight | AtomicWeight | | |
| Covalent radius | CovalentRad | | |
| Electronegativity | Electroneg | | |
| Volume (at 0K from DFT) | Volume | | |
| Melting point | Melting T | | |
| Number of valence electrons | NValence | | |
| Filled $s$ orbitals | NsValence | | |
| Filled $p$ orbitals | NpValence | | |
| Filled $d$ orbitals | NdValence | | |
| Unfilled valence orbitals | NUnfilled | | |
| Unfilled $s$ orbitals | NsUnfilled | | |
| Unfilled $p$ orbitals | NpUnfilled | | |
| Heat capacity at 800 K | Cp | | "Thermo" |
| Enthalpy at 800 K (relative to the enthalpy of the element in its stable state at 298 K) | H | | |
| Entropy at 800 K | S | | |
| Enthalpy of melting | Melting H | | |
| Entropy of melting | Melting S | | |
| Number of coexisting phases as extrapolated from binary descriptions | Extrapolated n_phase | Absolute value | "CALPHAD" |
| Gibbs energy as extrapolated from binary descriptions | Extrapolated Gibbs | | |
| Excess Gibbs energy in the liquid phase as extrapolated from binary descriptions | Extrapolated Ex G Liq | | |

## 2.3. Machine learning classification model

### 2.3.1. Random forest classifier

The random forest classifier (RFC) [18] is an ensemble learning algorithm in which a large number of decision trees are individually trained using random subsamples of the dataset. For an input sample, the probability of each class is calculated by averaging the votes of the trees. One advantage of using RFCs is that they provide a measure of the relative importance of each descriptor to the prediction model, thereby facilitating a straightforward feature selection. Besides, RFCs naturally provide probability estimates for each class. In this study, the RFC is implemented using the Scikit-learn library [20]. In the RFC, there are two important hyperparameters: the number of trees in the forest (n_estimators) and the fraction of samples randomly drawn from the dataset to train each tree (max_samples). Thus, these are tuned by following the procedure described in Section 2.3.2.

### 2.3.2. Model optimization and performance evaluation

The ML workflow used in the present study is schematized in Fig. 2 and detailed below.

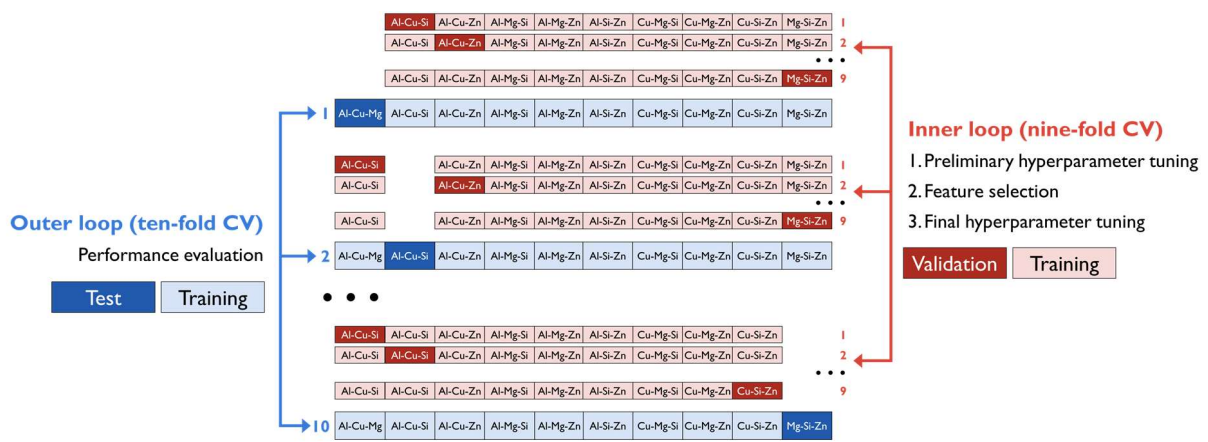

Fig. 2 – Schematic of the nested approach for optimizing the model and evaluating its performance. CV stands for cross-validation.

To evaluate the performance of the models, a 10-fold cross-validation is used: each of the 10 ternaries of the Al-Cu-Mg-Si-Zn system is in turn considered unknown, and its 800 K isothermal section is predicted using the classification model trained by the 9 other ternaries. The optimization of each classification model is performed by non-random nine-fold cross-validation; that is, each of the nine known ternaries is in turn taken as the validation set, while the remaining eight others form the training set. During this stage, hyperparameters are tuned, and features are selected such that the prediction accuracy is maximized following a three-step process.

1. A preliminary hyperparameter tuning is performed using a grid search approach in which "n_estimators" is varied from 50 to 200 in steps of 50, and "max_samples" from 0.2 to 0.7 in steps of 0.1. All the other parameters are set to default values in Scikit-learn 1.0.2 [20].

2. Feature selection is performed while keeping the hyperparameters fixed at the values selected in the previous step. (i) Feature importance is calculated across the nine-fold cross-validations, and on this basis, the least important features are eliminated. (ii) Using this smaller feature set, the classification models are trained, and step (i) is repeated. Step (ii) is repeated over several iterations until the set of features that maximizes the prediction accuracy is obtained. For more information on the number of features eliminated at each iteration, the example provided in Supplementary Note B (Fig. S1(c)) may be referred to.

3. The final hyperparameter tuning is performed based on the optimal set of features using a grid search approach. "n_estimators" is varied from 100 to 400 in steps of 100, and "max_samples" is varied from 0.2 to 0.8 in steps of 0.1. All the other parameters are set to default values in Scikit-learn [20].

A classification model with a high predictive potential is thus obtained from the nine ternaries and applied to the prediction of the phase diagram that is considered unknown. A practical example of the model optimization procedure can be referred to in Supplementary Note B.

# 3. Results

## 3.1. Classification performance

The average of prediction accuracies realized using RFCs across the 10 considered ternary sections is summarized in Fig. 3(a). It is calculated from the confusion matrices presented in Supplementary Note A (Tables S2-4). It is noted that this metric corresponds to the accuracy calculated using nested cross-validation [42], which is also called double cross-validation (Fig. 2). Here, we compare the results obtained using three different feature sets: "Magpie," "Magpie + Thermo," and "Magpie + Thermo + CALPHAD." In addition, the results obtained when extrapolating the ternaries from their constitutive binaries using CALPHAD calculations are also presented for comparison. It is emphasized that, in this study, a given A-B-C ternary isothermal section comprises 1326 discretized points, 150 of which are located on the A-B, A-C, and B-C binary edges. When a section is predicted, the points located on its binary edges are included in the test set, but they are already observed during training as they are also part of other known similar systems. If only the points inside the predicted ternary section are considered when evaluating the performance, i.e., if the 150 points located on the binary edges are eliminated from the test set, the average accuracies of the RFCs decrease from 0.62, 0.64, and 0.84 to 0.57, 0.60, and 0.82, respectively, for the three descriptor sets considered.

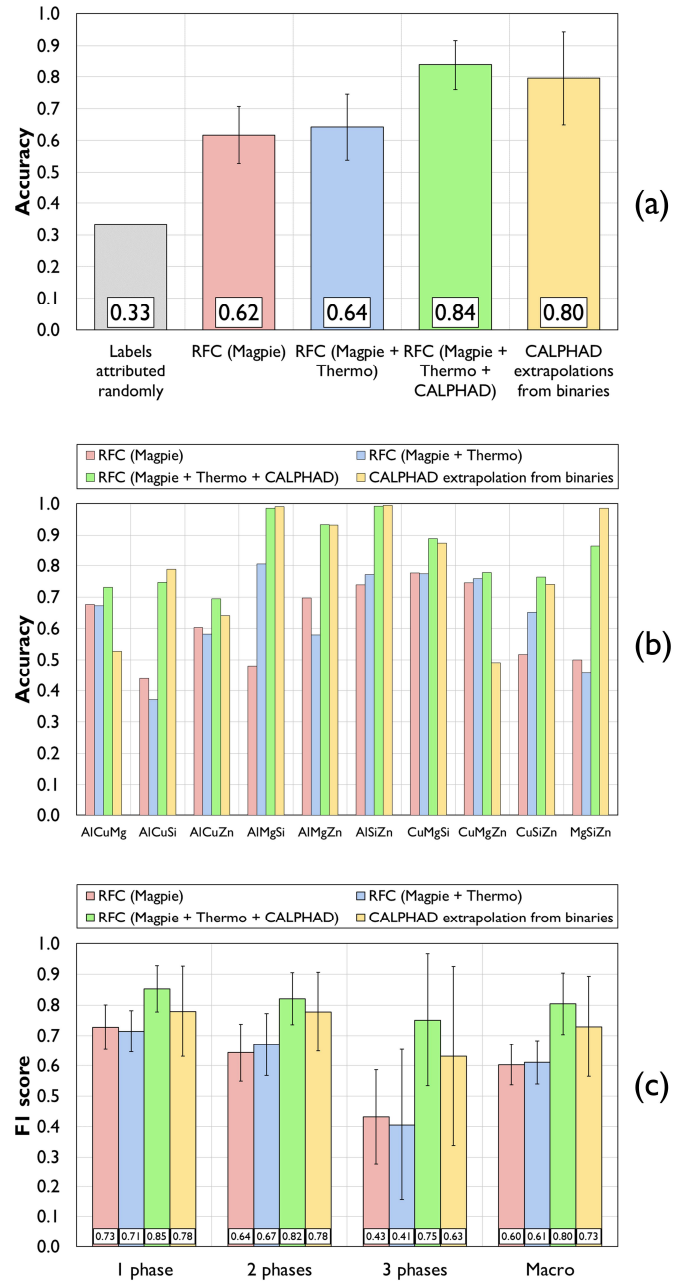

Fig. 3 – Classification performance of each model compared with CALPHAD extrapolations. (a) Average of prediction accuracies for the 10 isothermal sections. (b) Accuracy for each individual case. (c) Average of F1 scores for each class and of macro F1 scores across the 10 sections. In (a) and (c), the error bars represent the 95% confidence interval calculated from the 10 individual cases.

Firstly, each of the investigated RFCs significantly outperformed the benchmark case, for which classes were attributed randomly, as presented in Fig. 3(a). This is an evidence that the prediction strategy used in this study is working as expected.

Although the prediction accuracy is slightly increased, the "Magpie + Thermo" model does not significantly outperform the model trained using "Magpie" features only. However, as is presented shortly, "Thermo" descriptors rank high in the calculated feature importance, which implies that these descriptors must be important for predicting phase diagrams.

It is also noted that the isothermal sections obtained using CALPHAD extrapolations from the binaries are on average more accurate than the ones predicted based on the "Magpie" and "Magpie + Thermo" descriptor sets. However, it is emphasized that CALPHAD extrapolations and RFC predictions are fundamentally different; each approach has its own strengths, as is discussed in Section 4. Moreover, these first two RFCs result in relatively accurate predictions based only on the composition and properties of the pure elements, whereas additional information is required to perform CALPHAD extrapolations.

The ternary isotherms are on average best predicted by the RFC trained using the "Magpie + Thermo + CALPHAD" descriptor set. A 20% increase in the accuracy is observed as compared to the two other investigated models, i.e., "Magpie" and "Magpie + Thermo." This significant improvement resulted mainly from the addition of only three features. Among them, the number of coexisting phases calculated by extrapolation from the binaries ("Extrapolated n_phase") can be considered to be an estimate of the target variable. Therefore, the last RFC can be considered as a simplified case wherein the error from the CALPHAD extrapolations must be predicted rather than the number of coexisting phases. This is similar to the so-called crude estimation of property strategy used in ML [43,44]. It is emphasized that, as shown in Fig. 3(b), when the CALPHAD extrapolations from the binaries are accurate, as in the Al-Mg-Si case, the results of the "Magpie + Thermo + CALPHAD" RFC are equally good. However, when they are inaccurate, as in the Al-Cu-Mg case, this RFC still performs better than the other two RFCs. The only case wherein CALPHAD extrapolations are found to be significantly more accurate than the "Magpie + Thermo + CALPHAD" RFC is the Mg-Si-Zn system. For this ternary, the probabilities of having two- or three-phase domains are both calculated to be close to 50% in the central part of its 800 K section (Supplementary Note D, Fig. S13). As a result, the predicted section is less accurate, although a precise understanding of the phase equilibria is still obtained by observing the probability distribution of each label. It is noted that, in addition to the RFC, we also use the SVM classifier [19] for these prediction problems, and slightly lower performances are obtained (Supplementary Note C).

Figure 3(c) summarizes the average of F1 scores for each class and of macro F1 scores across the 10 considered ternary sections. It should be noted that, for each of the three RFCs, the F1 score of the "three-phase domain" class is undefined for two systems out of the 10 systems examined, wherein this outcome is never predicted. Thus, these cases are omitted from the calculation of the mean F1 score, which possibly results in an optimistically biased performance evaluation of this class. As shown in Fig. 3(c), the prediction of three-phase domains is the most difficult task. For this class, we confirmed that the precision of the RFCs is generally relatively high, but their recall is low. This indicates that insufficient points are labeled as three-phase domains in the predicted phase diagrams. In contrast, the single-phase domains are best predicted. This is especially true when using the "Magpie + Thermo + CALPHAD" descriptor set, wherein an average F1 score of 0.85 is obtained. This is a better performance than those of both CALPHAD extrapolations and other RFCs. We confirmed that adoption as a feature of the excess Gibbs energy in the liquid phase, as extrapolated from the binaries, played an important role in achieving this outcome. It is interesting to discuss this positive result in the context of the prediction of single-phase solid solution for multi-principal element alloys (MPEAs, also referred to as high-entropy alloys). In related studies, the enthalpy of mixing is often considered as a descriptor [45–50]. However, it is calculated based on Miedema's scheme with the use of enthalpy data for liquids [51], and it was noted [51] that this treatment generally results in relatively large errors. Furthermore, only the configurational entropy is usually considered. It has been highlighted [51,52] that this ideal behavior is more of an exception than the norm in MPEAs. Therefore, it is expected that an enhanced performance can be realized by considering the excess thermodynamic properties as calculated using the CALPHAD framework. It is expected that a reliable thermodynamic description of these complex multicomponent solutions can be obtained via extrapolation from binaries and ternaries, as suggested by a recent study on the enthalpy of mixing of the Co-Cr-Fe-Mn-Ni FCC phase [53]. It is emphasized that both the mixing enthalpy and entropy can be calculated in a spreadsheet directly from the composition and interaction parameters found in published thermodynamic databases. For materials informatics studies wherein a large compositional space is investigated, simply considering extrapolations from binaries may be an interesting compromise. Because this is a nontrivial task for those unfamiliar with the CALPHAD formalism, the calculation of the excess thermodynamic properties by extrapolation from the binaries is detailed in the appendix. Furthermore, a Python script is made available to obtain these easily (Section 2.2.3).

## 3.2. Important features

For the present classification task, we consider feature selection to be a critical part of the model optimization. Indeed, before and after this stage, which is described as step 2 in Section 2.2.2, a flat increase in accuracy of 8.4% is observed on average. In comparison, hyperparameter tuning performed during steps 1 and 3 only resulted in variations in accuracy of approximately 1%. After the feature selection, 11 features remained on average without any significant shift for the three descriptor sets.

Figure 4 presents the selected features for each descriptor set. Only the features that are selected in at least two of the 10 cases are presented herein. On comparing the results obtained using the "Magpie" and "Magpie + Thermo" descriptor sets, it is found that the selected features differ significantly, although the prediction accuracies are close (Fig. 2). Indeed, in the case of "Magpie + Thermo," the thermodynamic properties tend to be selected as important features. These "Thermo" descriptors are also actively selected with the use of the "Magpie + Thermo + CALPHAD" descriptor set. This implies that these thermodynamic descriptors are useful for predicting phase diagrams. For the RFC trained using the "Magpie + Thermo + CALPHAD" set, the number of coexisting phases calculated by extrapolation from the binaries is the most important feature. Another noteworthy feature is the excess Gibbs energy in the liquid phase as extrapolated from the binaries. Indeed, this descriptor is selected in all 10 cases and ranked third on average in terms of feature importance.

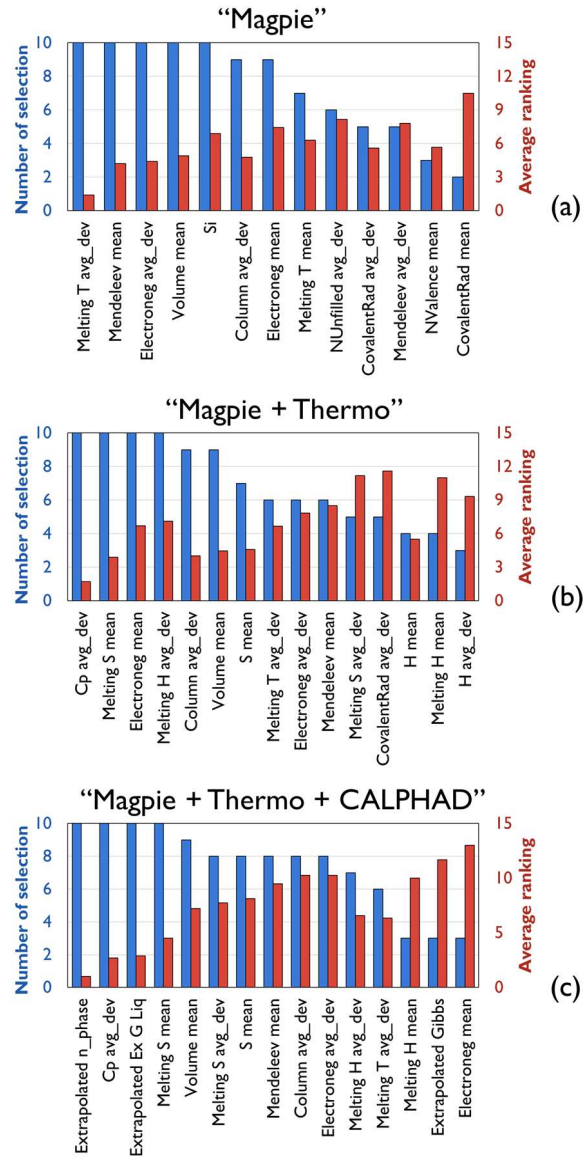

Fig. 4 – Overview of feature selection when using (a) "Magpie," (b) "Magpie + Thermo," and (c) "Magpie + Thermo + CALPHAD" descriptor sets. The average number of selection and average ranking are summarized.

### 3.3. Use of predictions to guide experiments

These predictions provide insights into phase equilibria and can thus be useful for planning experiments. To illustrate this, the Cu-Mg-Zn ternary system is considered as an example. This system is rather complicated, notably owing to the formation of the ternary Laves C36 phase and the significant solubility of Zn in the $Cu_2Mg$ (Laves C15) phase. As a result, the 800 K isothermal section presented in Fig. 5 (a) differs significantly from that extrapolated from the binaries presented in Fig. 5 (b). As discussed in Section 3.1, the RFC tends to underestimate the probability of having three-phase domains. As a result, this class is almost absent from the section predicted using the "Magpie + Thermo + CALPHAD" descriptor set presented in Fig. 5 (c). In contrast, the prediction of single-phase domains is significantly better.

Figures 5 (d)–(f) present the predicted probabilities of having single-, two-, and three-phase domains, respectively. Although the section extrapolated from the binaries is characterized by a significantly extended single-phase liquid domain (Fig. 5(b)), this behavior is not predicted by the RFC, as shown in Fig. 5(d). Furthermore, from Figs. 5(e) and (f), it appears that the predicted probabilities for two- or three-phase domains are large in the central part of the phase diagram. This is consistent with the real isothermal section presented in Fig. 5 (a). A preliminary understanding of the phase equilibria can thus be obtained from the predictions via a subjective analysis. For example, if the objective is to find three-phase domains in the phase diagram, the candidate points can be selected using the probability presented in Fig. 5(f), i.e., the dark-red region is a promising area. These areas are generally found to be consistent with the three-phase domains in real isothermal sections.

If both the extrapolated section obtained via CALPHAD calculations and the predicted section are available, the differences between these provide useful information for guiding experiments in a more objective manner. In Fig. 6, the probability of having another class than the one obtained via CALPHAD extrapolations is displayed at each investigated point of the Cu-Mg-Zn section. The dark areas are where the extrapolations are likely to be incorrect according to the model, possibly owing to the formation of ternary phases. These areas are thus where experiments should be conducted first. This strategy resembles the uncertainty sampling approach for constructing phase diagrams efficiently proposed in [14–17].

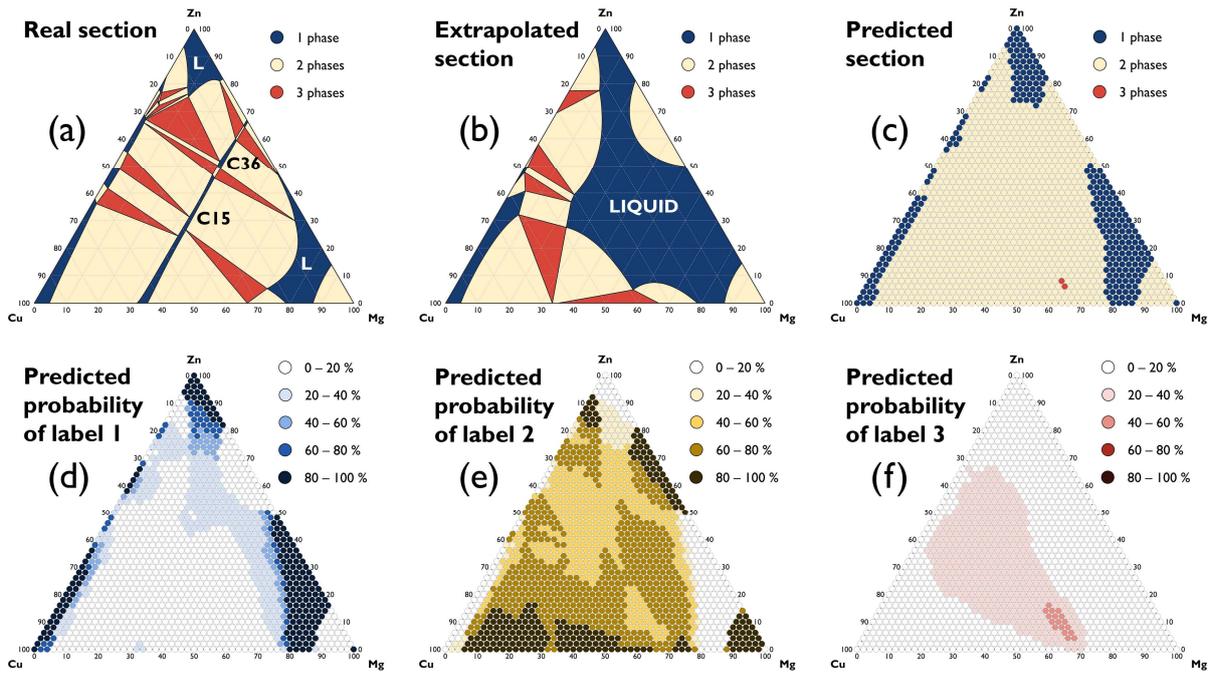

Fig. 5 – 800 K isothermal section of the Cu-Mg-Zn as (a) calculated based on the assessment from Ref. [29] (real section), (b) extrapolated from descriptions of the Cu-Mg [36], Cu-Zn [40], and Mg-Zn [27] binaries by CALPHAD calculations, and (c) predicted by RFC using the "Magpie + Thermo + CALPHAD" descriptor set. (d), (e), and (f) present probabilities predicted by RFC of having single-, two-, and three-phase domains, respectively.

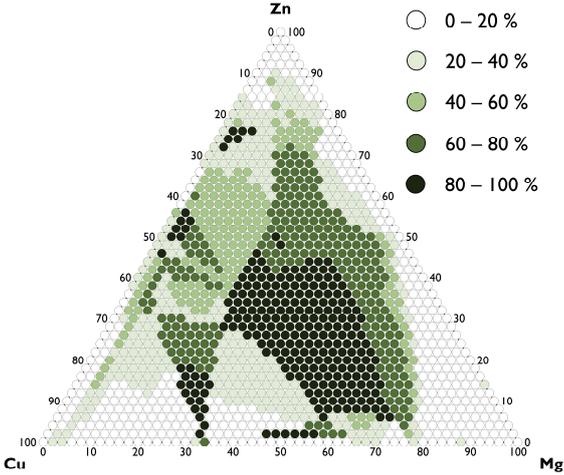

Fig. 6 – Probabilities across the Cu-Mg-Zn 800 K isothermal section of having another class than that obtained via CALPHAD extrapolations as predicted by RFC using the "Magpie + Thermo + CALPHAD" descriptor set.

# 4. Discussion

The results obtained for each of the 10 isothermal sections and three descriptor sets considered are summarized in Supplementary Note D. In general, predictions by RFCs are useful in cases wherein the phase diagram features an appreciable proportion of single-phase domains, as in the Cu-Mg-Zn case discussed in Section 3.2 as well as the Al-Cu-Mg and Al-Cu-Zn systems. In contrast, limited information is obtained in the cases wherein three-phase domains are largely predominant, such as in the case of the Al-Cu-Si and Cu-Mg-Si systems. Therefore, a better performance is expected at relatively high temperatures close to the liquidus, where solutions extend into the diagram, but some compounds remain stable. Furthermore, the best prediction results are obtained when using the "Magpie + Thermo + CALPHAD" descriptor set for the Al-Mg-Si, Al-Si-Zn, and Mg-Si-Zn systems. In all three cases characterized by the absence of ternary phases, the 800 K isothermal section is predicted not only with a high level of accuracy, but also with a high level of confidence.

It is emphasized that CALPHAD extrapolations and the present ML predictions are different in nature. In CALPHAD, phase diagrams are calculated from the thermodynamic description of their constitutive subsystems. The CALPHAD method is successful in extrapolating quaternary and higher-order phase diagrams from binaries and ternaries. However, extrapolations of ternaries from binaries are less consistent, and a binary cannot be reliably extrapolated from unaries. An important limitation of CALPHAD extrapolations is that a compound, or the solubility of an element in a phase, cannot be predicted if it does not already appear in the subsystems. In the present ML approach, phase diagrams are predicted by analogy with other systems. In contrast to CALPHAD extrapolations, ML can provide indications of the presence of new phases, as discussed in Section 3.3. Moreover, a measure of confidence in the predictions can be obtained from the class probability distribution.

In the present dataset, the classes are relatively well-balanced. However, for other systems and temperatures, class imbalance can become a problem, in which case undersampling and oversampling techniques may be important in achieving good results.

This study is focused on cases wherein the phase diagram of interest is completely unknown, except for its constitutive binaries. If partial data are available in the ternary phase diagram, they can be directly included in the training set. The strategies for guiding the experiments discussed herein can thus become part of an iterative process, such as black-box optimization [54,55], wherein the newly obtained data are used to refine the predictions before planning for the next step. It would be interesting to investigate the extent to which the performance of the model could be improved at each iteration.

We should emphasize that this study is our first attempt at predicting unknown phase diagrams from known phase diagrams. A certain degree of success is achieved, although only nine phase diagrams are used to train the models, and the prediction target is the number of coexisting phases. Future perspectives for improving the present approach include increasing the size of the dataset, introducing other appropriate labels for each phase, and accounting for temperature variations. To achieve this, the phase diagram data for various systems must be collected by performing CALPHAD calculations. These potential research directions are being considered for future works.

## Conclusions

Phase diagram prediction using ML was investigated by focusing on the ternary subsystems of the Al-Cu-Mg-Si-Zn quinary. Each of the 10 ternaries was considered to be unknown, and the number of coexisting phases across its 800 K isothermal section was predicted from the 9 others. Special emphasis was placed on introducing new descriptors for predicting the phase diagrams. It is emphasized that impactful features were obtained from CALPHAD extrapolations from the binary systems. To the best of our knowledge, this is a new strategy for generating descriptors, and it significantly improved the model performance in the present study. Using the most complete set of descriptors and a random forest algorithm, an average accuracy of 84% with a standard deviation of 11% was obtained across the 10 considered ternary sections. It should be noted that the accuracy is dependent on the system, temperature, and training set. It was demonstrated that the predictions can provide insights into phase equilibria in unknown materials systems, especially when combined with CALPHAD extrapolations. It is thus concluded that the proposed approach is promising for promoting more efficient determination of phase diagrams. We are currently working on the construction of a larger dataset to expand our approach to other elements.

# Acknowledgements

This study was supported by a project subsidized by the Core Research for Evolutional Science and Technology (CREST) program of the Japan Science and Technology Agency (JST) (Grant No. JPMJCR17J2). We thank Masanori Enoki for sharing his thermodynamic databases.

# Data availability

All thermodynamic databases used in the calculations are available in the NIMS Computational Phase Diagram Database (CPDDB) [56]. The complete dataset used in this study is shared as supplementary material.

# CRediT authorship contribution statement

Guillaume Deffrennes: Data Curation, Formal Analysis, Investigation, Methodology, Visualization, Writing – Original Draft Preparation. Kei Terayama: Methodology, Writing – Review & Editing. Taichi Abe: Data Curation, Writing – Review & Editing. Ryo Tamura: Conceptualization, Funding Acquisition, Methodology, Supervision, Writing – Original Draft Preparation.

# Competing interests

The authors declare no competing interests.

# Appendix. CALPHAD extrapolation of the excess thermodynamic properties of a substitutional solution from its binary subsystems

For a given A-B binary substitutional solution denoted by $\varphi$, the molar excess Gibbs energy $^{ex}G$ is modeled within the CALPHAD framework using a Redlich–Kister polynomial that includes interaction parameters:

$$^{ex}G^\varphi = x_A x_B \sum_v (\,^v a_{A,B}^\varphi + \,^v b_{A,B}^\varphi T)(x_A - x_B)^v, \tag{1}$$

with $x_i$ the molar fraction of constituent $i$, and $v$ an integer. $^v a_{i,j}^\varphi$ and $^v b_{A,B}^\varphi$ form the interaction parameters of order $v$, which can be obtained from CALPHAD publications in tables under the notation $^v L_{i,j}^\varphi$. In thermodynamic databases in TDB format such as those available in NIMS CPDDB [56], binary interaction parameters are found after the keyword "PARAMETER $L(\varphi, i, j; v)$".

For a substitutional solution of $n$ constituents ($n > 2$), the main contribution to the excess Gibbs energy is obtained by extrapolation from its binary subsystems. Various extrapolation methods are available, as discussed in [57,58]. The Muggianu formalism is most commonly used because the contribution from each binary can be directly calculated using Eq. (1), without the requirement for any weighting. Thus, the contribution to the excess thermodynamic properties from the binaries is obtained by the summation of Eq. (1) across all the cases as follows:

$$^{bin,ex}G^\varphi = \sum_i \sum_j x_i x_j \sum_v (\,^v a_{i,j}^\varphi + \,^v b_{i,j}^\varphi T)(x_i - x_j)^v, \tag{2}$$

$$^{bin,ex}H^\varphi = \sum_i \sum_j x_i x_j \sum_v \,^v a_{i,j}^\varphi (x_i - x_j)^v, \tag{3}$$

$$^{\text{bin,ex}}S^\varphi = -\sum_i \sum_j x_i x_j \sum_v {}^v b_{i,j}^\varphi (x_i - x_j)^v, \tag{4}$$

where $^{\text{bin,ex}}H^\varphi$ and $^{\text{bin,ex}}S^\varphi$ are the excess enthalpy and excess entropy as extrapolated from the binaries, i.e., on disregarding ternary and eventual higher-order interactions between the elements. The excess enthalpy directly corresponds to the enthalpy of mixing, whereas the configurational entropy must be added to the excess entropy to obtain the entropy of mixing.

# Supplementary notes for

# A machine learning–based classification approach for phase diagram prediction

Guillaume Deffrennes[a,*], Kei Terayama[b], Taichi Abe[c,d], and Ryo Tamura[a,d,e,f,*]

[a] International Center for Materials Nanoarchitectonics (WPI-MANA), National Institute for Materials Science, 1-1 Namiki, Tsukuba, Ibaraki 305-0044, Japan

[b] Graduate School of Medical Life Science, Yokohama City University, 1-7-29, Suehiro-cho, Tsurumi-ku, Kanagawa 230-0045, Japan

[c] Research Center for Structural Materials, National Institute for Materials Science, 1-2-1 Sengen, Tsukuba, Ibaraki 305-0047, Japan

[d] Research and Services Division of Materials Data and Integrated System, National Institute for Materials Science, 1-1 Namiki, Tsukuba, Ibaraki 305-0044, Japan

[e] RIKEN Center for Advanced Intelligence Project, 1-4-1 Nihonbashi, Chuo-ku, Tokyo 103-0027, Japan

[f] Graduate School of Frontier Sciences, The University of Tokyo, 5-1-5 Kashiwa-no-ha, Kashiwa, Chiba 277-8561, Japan

∗ Corresponding authors at: National Institute for Materials Science, 1-1 Namiki, Tsukuba, Ibaraki 305-0044, Japan. E-mail addresses: DEFFRENNES.Guillaume@nims.go.jp and TAMURA.Ryo@nims.go.jp

**Supplementary Note A: Detailed distribution of classes, and confusion matrix for each random forest classifier**

The distribution of classes for each of the isothermal section considered is detailed in Table S1. The confusion matrices for each of the "Magpie," "Magpie + Thermo," and "Magpie + Thermo + CALPHAD" random forest classifiers (RFCs) are listed in Tables S2-S4.

**Table S1.** Distribution of classes for each of the 10 800 K isothermal sections considered

| System | 1. Single-phase domains | 2. Two-phase domains | 3. Three-phase domains |
|---|---|---|---|
| AlCuMg | 266 | 742 | 318 |
| AlCuSi | 65 | 400 | 861 |
| AlCuZn | 496 | 583 | 247 |
| AlMgSi | 36 | 389 | 901 |
| AlMgZn | 990 | 316 | 20 |
| AlSiZn | 55 | 1092 | 179 |
| CuMgSi | 19 | 340 | 967 |
| CuMgZn | 204 | 928 | 194 |
| CuSiZn | 101 | 835 | 390 |
| MgSiZn | 31 | 438 | 857 |
| Total | 2263 | 6063 | 4934 |

**Table S2.** Confusion matrix for each "Magpie" RFC

| System | 1 as 1 | 1 as 2 | 1 as 3 | 2 as 1 | 2 as 2 | 2 as 3 | 3 as 1 | 3 as 2 | 3 as 3 |
|---|---|---|---|---|---|---|---|---|---|
| AlCuMg | 248 | 18 | 0 | 111 | 615 | 16 | 21 | 264 | 33 |
| AlCuSi | 39 | 20 | 6 | 1 | 292 | 107 | 0 | 611 | 250 |
| AlCuZn | 222 | 272 | 2 | 3 | 577 | 3 | 0 | 247 | 0 |
| AlMgSi | 36 | 0 | 0 | 64 | 318 | 6 | 5 | 613 | 283 |
| AlMgZn | 614 | 376 | 0 | 1 | 309 | 6 | 0 | 20 | 0 |
| AlSiZn | 45 | 10 | 0 | 16 | 780 | 296 | 0 | 24 | 155 |
| CuMgSi | 15 | 3 | 1 | 9 | 316 | 15 | 1 | 267 | 699 |
| CuMgZn | 167 | 37 | 0 | 72 | 762 | 94 | 3 | 133 | 58 |
| CuSiZn | 84 | 11 | 6 | 5 | 445 | 385 | 0 | 235 | 155 |
| MgSiZn | 29 | 2 | 0 | 15 | 411 | 12 | 0 | 634 | 223 |
| Total | 1499 | 749 | 15 | 297 | 4825 | 940 | 30 | 3048 | 1856 |

**Table S3.** Confusion matrix for each "Magpie + Thermo" RFC

| System | 1 as 1 | 1 as 2 | 1 as 3 | 2 as 1 | 2 as 2 | 2 as 3 | 3 as 1 | 3 as 2 | 3 as 3 |
|---|---|---|---|---|---|---|---|---|---|
| AlCuMg | 171 | 95 | 0 | 28 | 713 | 1 | 1 | 310 | 7 |
| AlCuSi | 39 | 24 | 2 | 1 | 294 | 105 | 0 | 704 | 157 |
| AlCuZn | 212 | 284 | 0 | 22 | 560 | 1 | 2 | 245 | 0 |
| AlMgSi | 36 | 0 | 0 | 27 | 271 | 91 | 0 | 140 | 761 |
| AlMgZn | 467 | 523 | 0 | 8 | 301 | 7 | 0 | 19 | 1 |
| AlSiZn | 38 | 17 | 0 | 5 | 864 | 223 | 0 | 57 | 122 |
| CuMgSi | 12 | 6 | 1 | 10 | 318 | 12 | 1 | 270 | 696 |
| CuMgZn | 165 | 39 | 0 | 93 | 829 | 6 | 5 | 178 | 11 |
| CuSiZn | 78 | 23 | 0 | 0 | 679 | 156 | 0 | 282 | 108 |
| MgSiZn | 30 | 1 | 0 | 15 | 406 | 17 | 0 | 685 | 172 |
| Total | 1248 | 1012 | 3 | 209 | 5235 | 619 | 9 | 2890 | 2035 |

**Table S4.** Confusion matrix for each "Magpie + Thermo + CALPHAD" RFC

| System | 1 as 1 | 1 as 2 | 1 as 3 | 2 as 1 | 2 as 2 | 2 as 3 | 3 as 1 | 3 as 2 | 3 as 3 |
|---|---|---|---|---|---|---|---|---|---|
| AlCuMg | 234 | 31 | 1 | 51 | 670 | 21 | 5 | 248 | 65 |
| AlCuSi | 39 | 19 | 7 | 1 | 235 | 164 | 0 | 146 | 715 |
| AlCuZn | 329 | 161 | 6 | 30 | 527 | 26 | 6 | 176 | 65 |
| AlMgSi | 36 | 0 | 0 | 2 | 387 | 0 | 0 | 16 | 885 |
| AlMgZn | 976 | 14 | 0 | 55 | 261 | 0 | 9 | 11 | 0 |
| AlSiZn | 52 | 3 | 0 | 0 | 1090 | 2 | 0 | 5 | 174 |
| CuMgSi | 13 | 5 | 1 | 4 | 249 | 87 | 0 | 49 | 918 |
| CuMgZn | 176 | 28 | 0 | 75 | 853 | 0 | 4 | 187 | 3 |
| CuSiZn | 69 | 29 | 3 | 1 | 616 | 218 | 0 | 62 | 328 |
| MgSiZn | 29 | 2 | 0 | 0 | 430 | 8 | 0 | 167 | 690 |
| Total | 1953 | 292 | 18 | 219 | 5318 | 526 | 24 | 1067 | 3843 |

**Supplementary Note B: Practical example of the RFC optimization procedure**

The RFC optimization procedure is illustrated in Fig. S1 while considering the prediction of the Al-Cu-Zn system using the "Magpie + Thermo" descriptor set as an example.

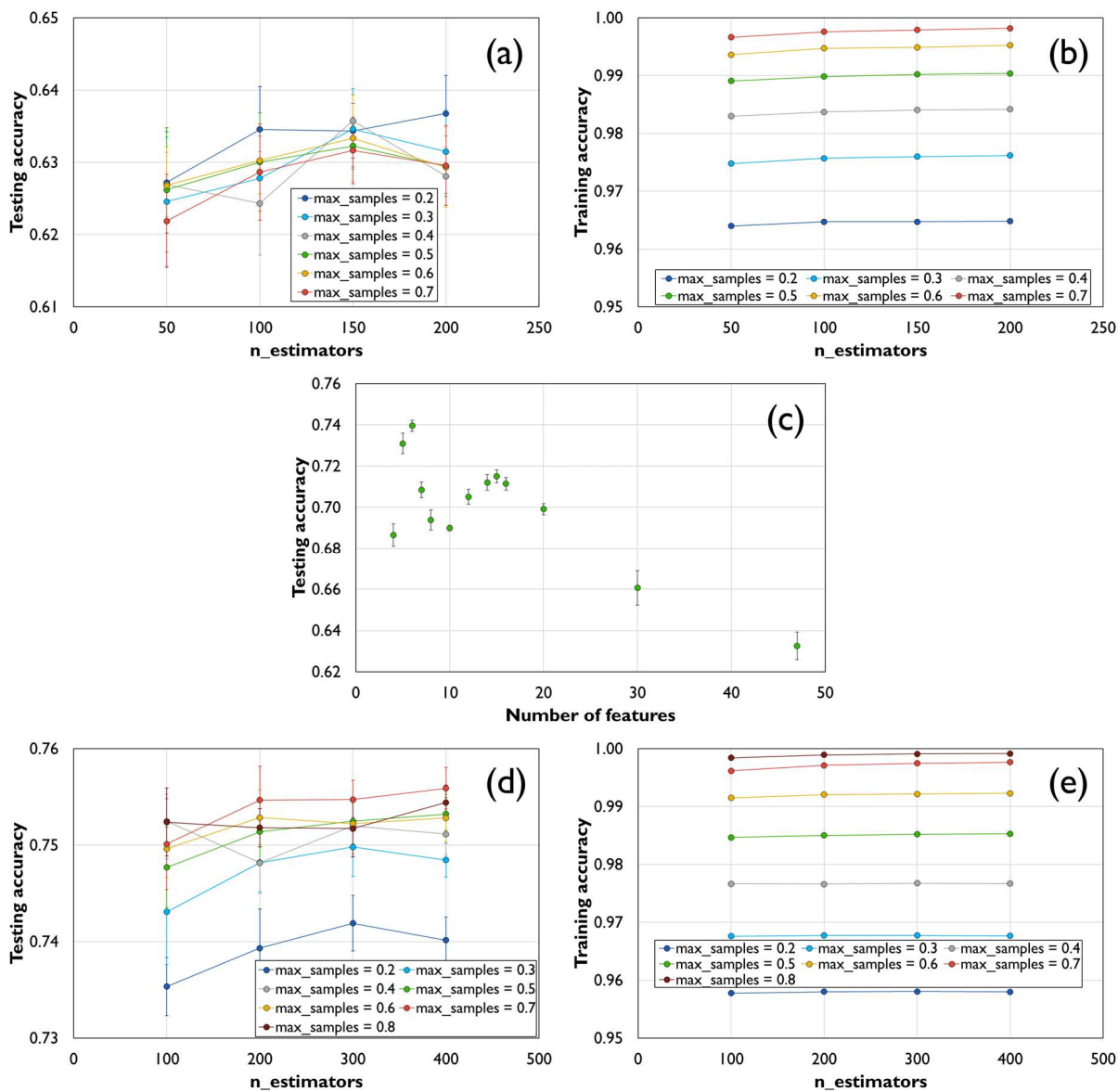

**Fig. S1.** RFC optimization procedure illustrated with the case wherein the Al-Cu-Zn system is considered unknown and the "Magpie + Thermo" descriptor set is used. The optimization is performed by non-random repeated nine-fold cross validation following a three-step process. (a-b) Preliminary hyperparameter tuning is performed using a grid-search approach. (c) Feature selection is performed while keeping the hyperparameters fixed at the values that maximized the prediction accuracy in the previous step (in this case, n_estimators = 200, and

max_samples = 0.2). Feature importance is calculated, some of the least important features are removed, and the process is repeated until the set of features that maximizes the prediction accuracy is determined. (d) and (e) Final hyperparameter tuning is performed using a grid-search approach. Each point in the subfigures represents the accuracy averaged across the nine folds, and the error bars represent the 95% confidence interval calculated from 8 to 12 repetitions.

**Supplementary Note C: Phase diagram prediction using SVM**

In addition to random forest, phase diagram prediction was also investigated using support vector machine (SVM) while focusing on the case wherein the "Magpie + Thermo + CALPHAD" dataset is used. The C-support vector classifier (SVC) was implemented using the Scikit-learn library [1]. The model was optimized following a procedure similar to that described in Section 2.3.2 of the manuscript for the RFC; however, the feature selection was not performed as the strategy used was not transposable to the SVM case. An "rbf" kernel was selected. Only the gamma and C hyperparameters were tuned in the inner cross-validation loop. This optimization procedure is illustrated in Fig. S2 while using the prediction of the Al-Cu-Zn system as an example.

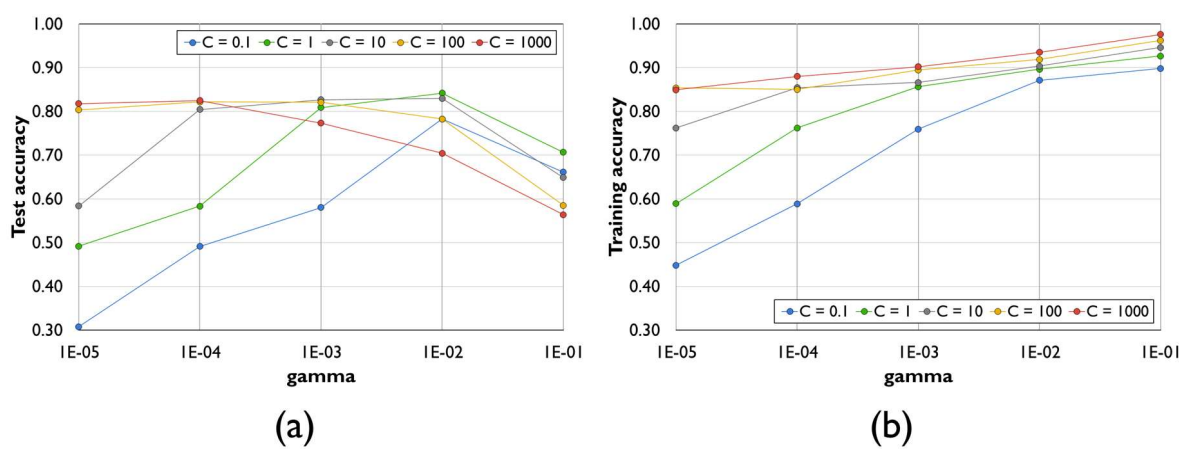

**Fig. S2.** SVM optimization procedure illustrated with the case wherein the Al-Cu-Zn system is considered unknown and the "Magpie + Thermo + CALPHAD" descriptor set is used. (a) Test accuracy. (b) Training accuracy.

The performances obtained using SVM and RFC are compared with the results of the CALPHAD extrapolations from the binaries in Fig. S3. The RFC performed slightly better than the SVM classifier. It is noted that an average of the training accuracies of 0.85 was obtained with SVM for the 10 ternaries, whereas the RFC provides a significantly better score of 0.96.

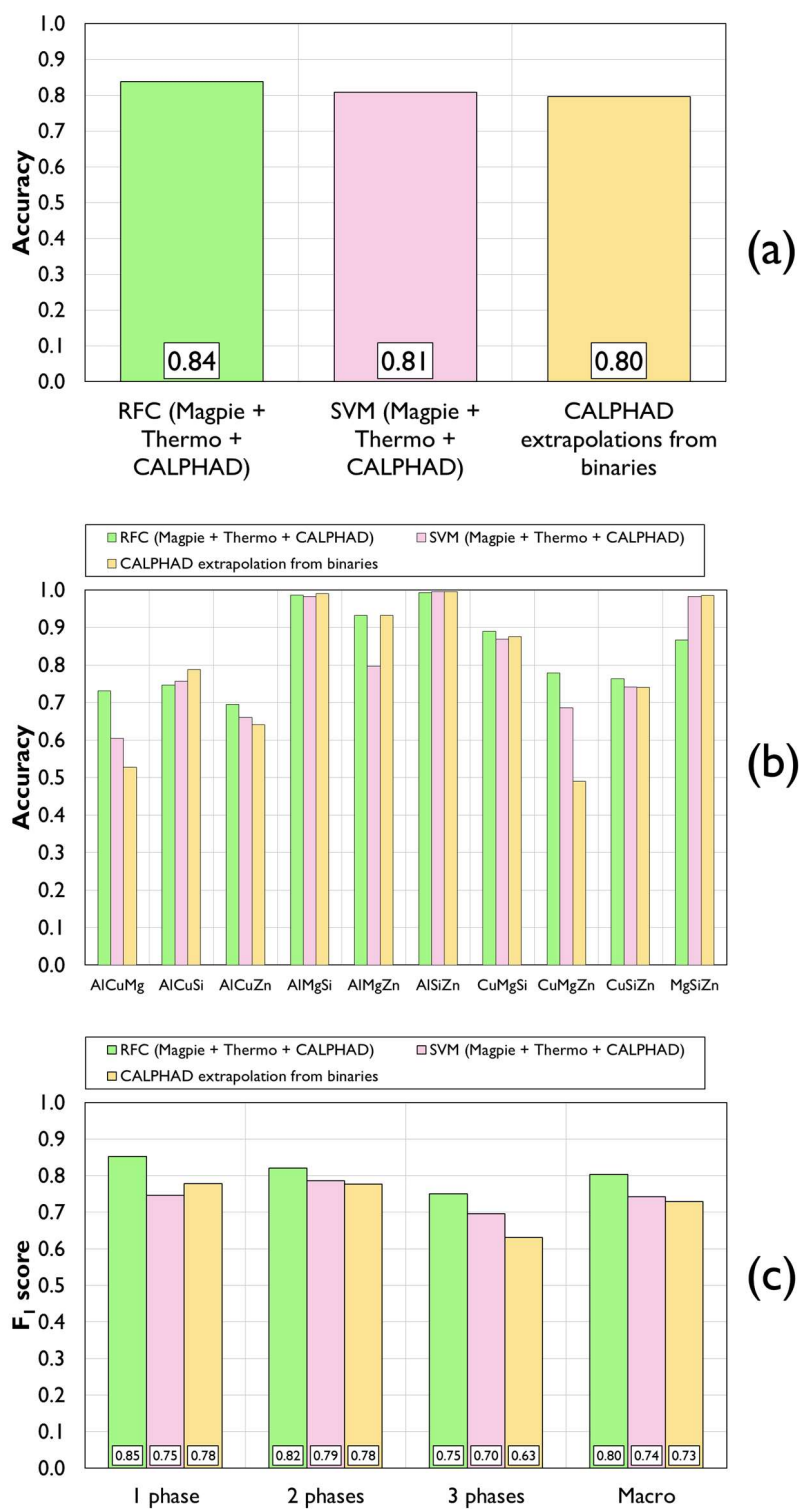

**Fig. S3.** Classification performance of the SVM model compared with the results obtained using RFC and CALPHAD extrapolations. (a) Average of prediction accuracies for the 10 ternaries. (b) Accuracy for each individual case. (c) Average of F1 scores for each class and of macro F1 scores across the 10 sections.

**Supplementary Note D: Detailed results of each prediction case**

For each of the ten systems and three descriptor sets considered, the results of the phase diagram prediction using RFCs are presented in Figs. S4–S13. From top to bottom, the subfigures correspond to the following:

- 1st row : (Left) Section calculated using the selected thermodynamic database (Section 2.1. of the manuscript) and (right) section calculated using extrapolation from the selected binary descriptions (Section 2.2.3. of the manuscript).

- 2nd row : (From left to right) Sections predicted using the "Magpie," "Magpie + Thermo," and "Magpie + Thermo + CALPHAD" descriptor sets, respectively.

- 3rd row : (From left to right) Probabilities across the section of having a single-phase domain as predicted using the "Magpie," "Magpie + Thermo," and "Magpie + Thermo + CALPHAD" descriptor sets, respectively.

- 4th row : (From left to right) Probabilities across the section of having a two-phase domain as predicted using the "Magpie," "Magpie + Thermo," and "Magpie + Thermo + CALPHAD" descriptor sets, respectively.

- 5th row : (From left to right) Probabilities across the section of having a three-phase domain as predicted using the "Magpie," "Magpie + Thermo," and "Magpie + Thermo + CALPHAD" descriptor sets, respectively.

- 6th row : (From left to right) Probabilities across the section of having another class than the one obtained using CALPHAD extrapolation as predicted using the "Magpie," "Magpie + Thermo," and "Magpie + Thermo + CALPHAD" descriptor sets, respectively.

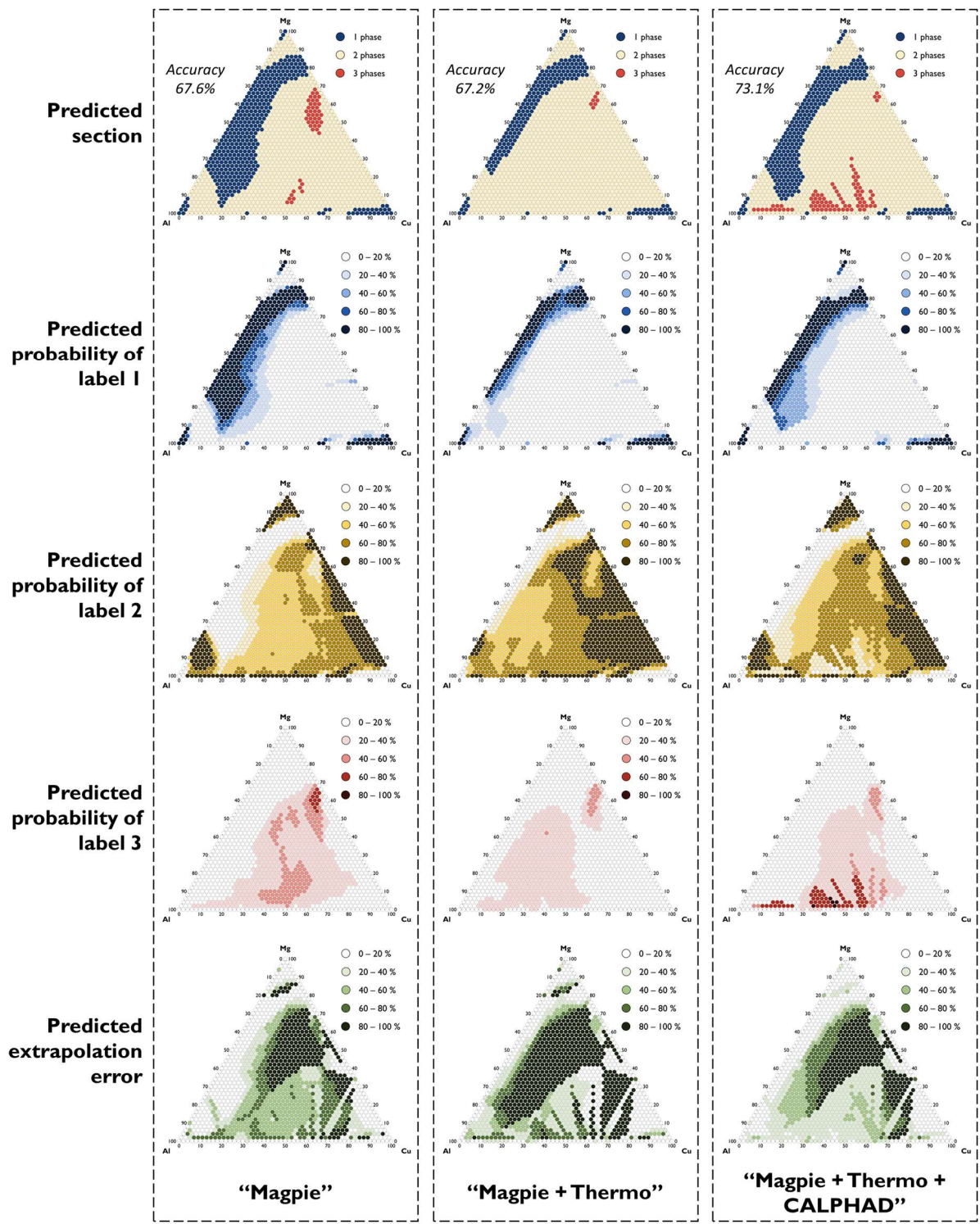

**Fig. S4.** Prediction of the isothermal section of the Al-Cu-Mg system at 800 K.

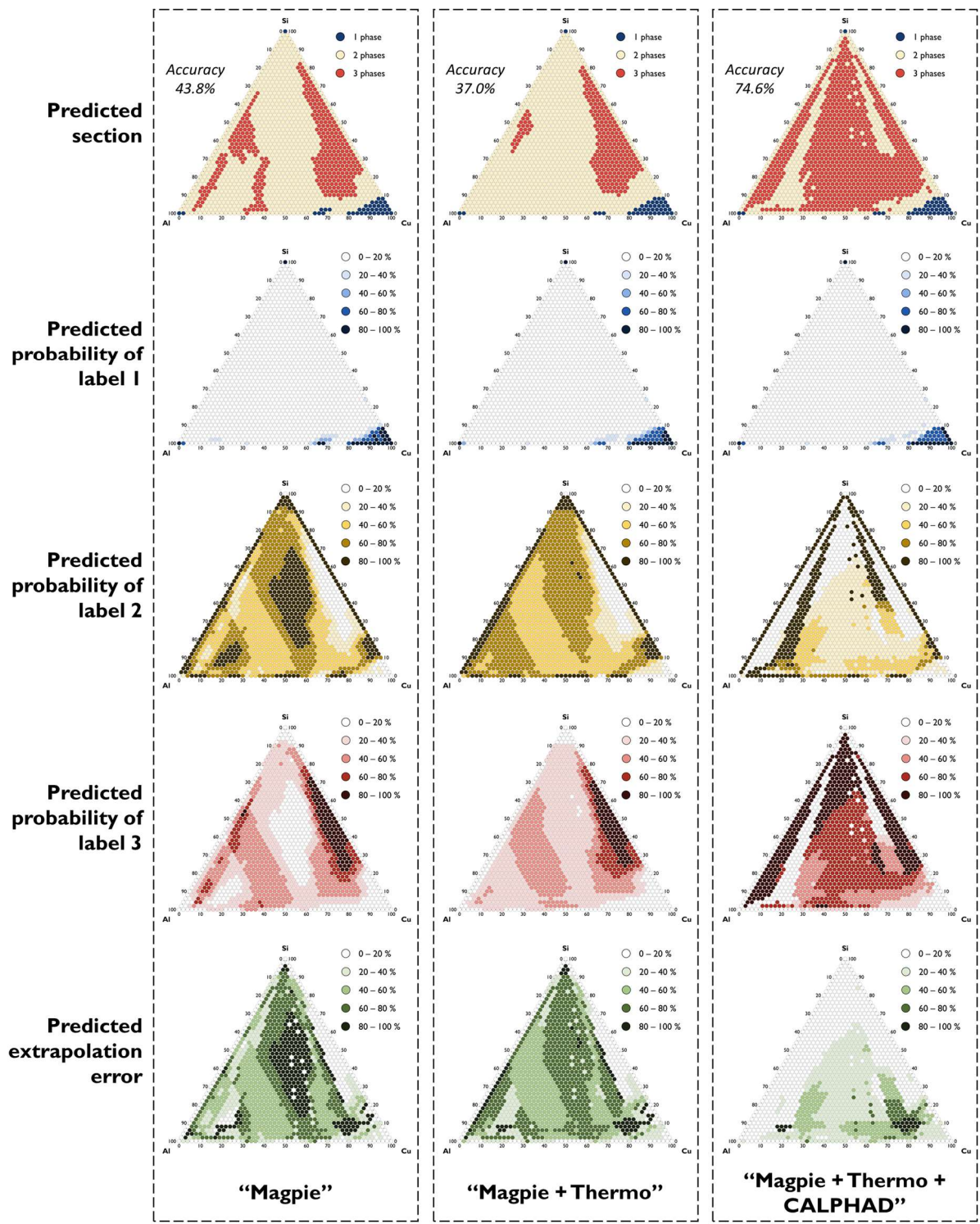

**Fig. S5.** Prediction of the isothermal section of the Al-Cu-Si system at 800 K.

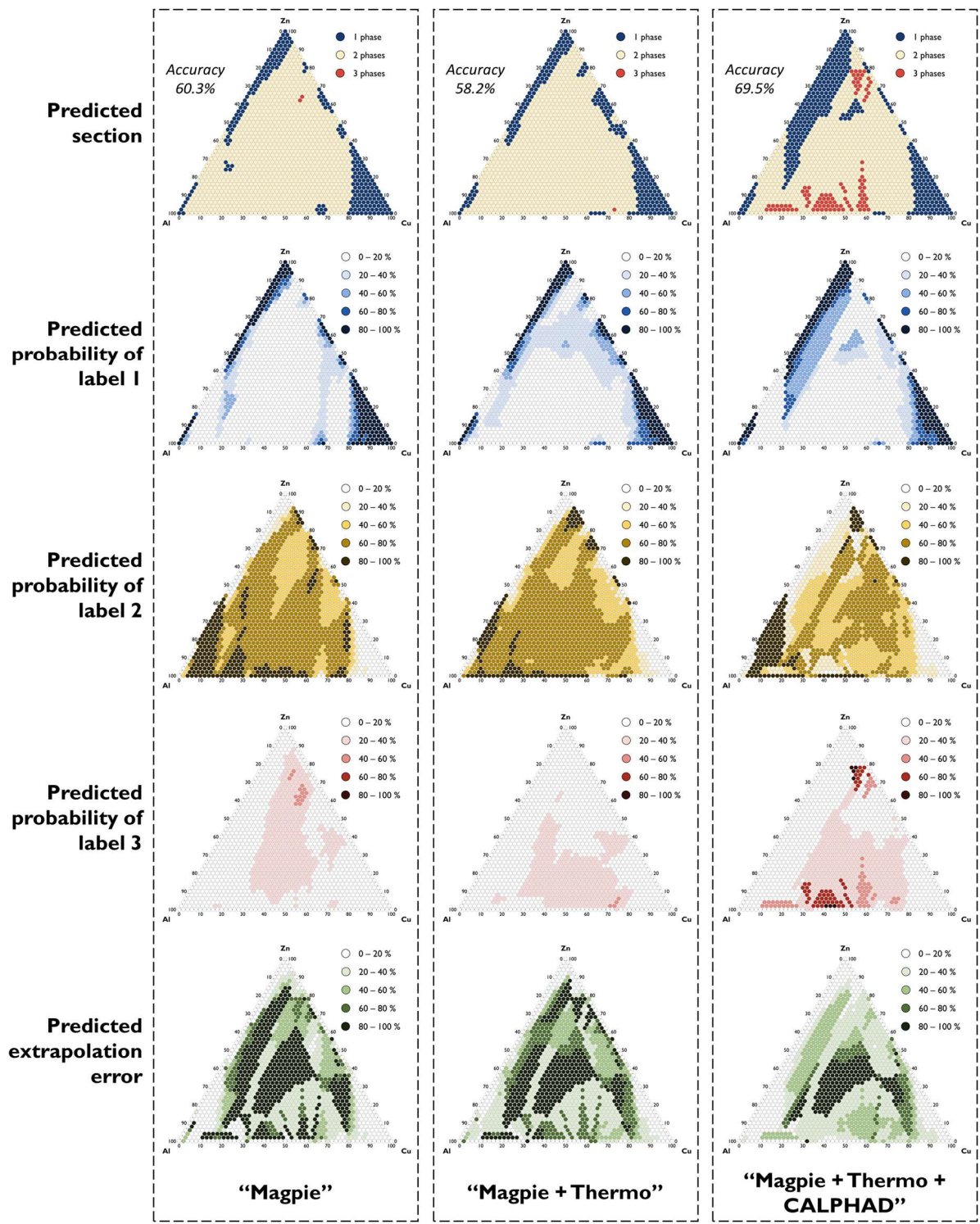

**Fig. S6.** Prediction of the isothermal section of the Al-Cu-Zn system at 800 K.

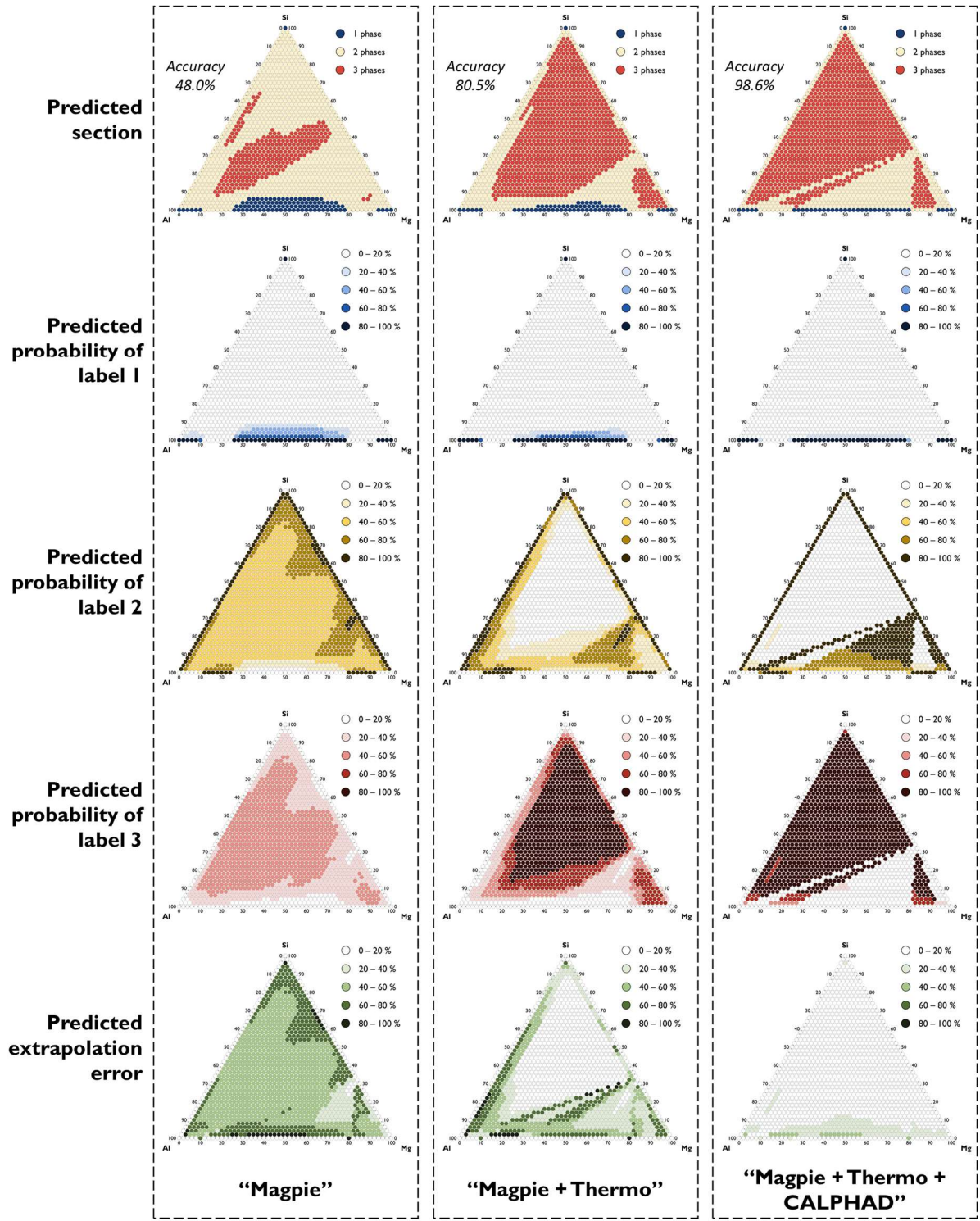

**Fig. S7.** Prediction of the isothermal section of the Al-Mg-Si system at 800 K.

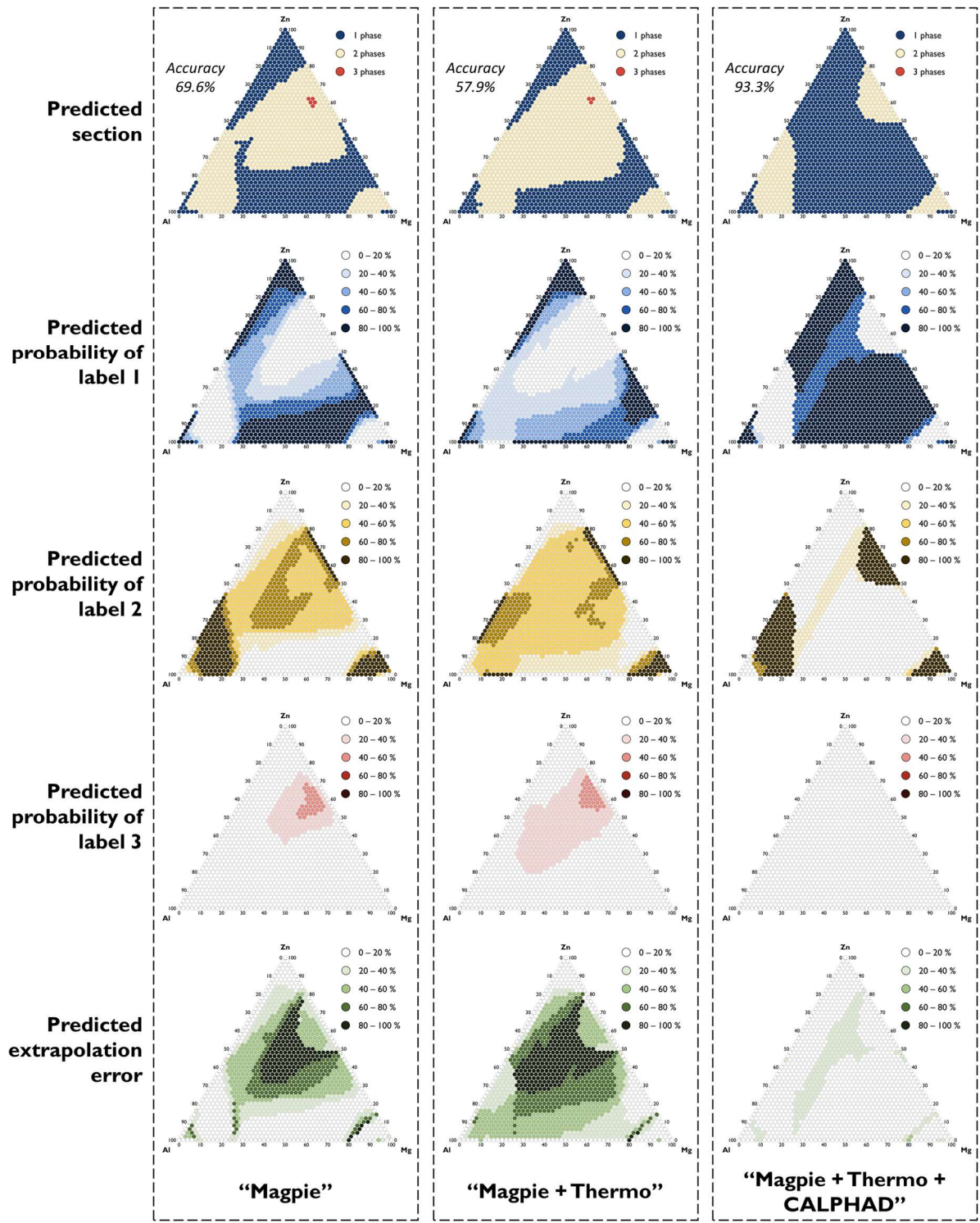

**Fig. S8.** Prediction of the isothermal section of the Al-Mg-Zn system at 800 K.

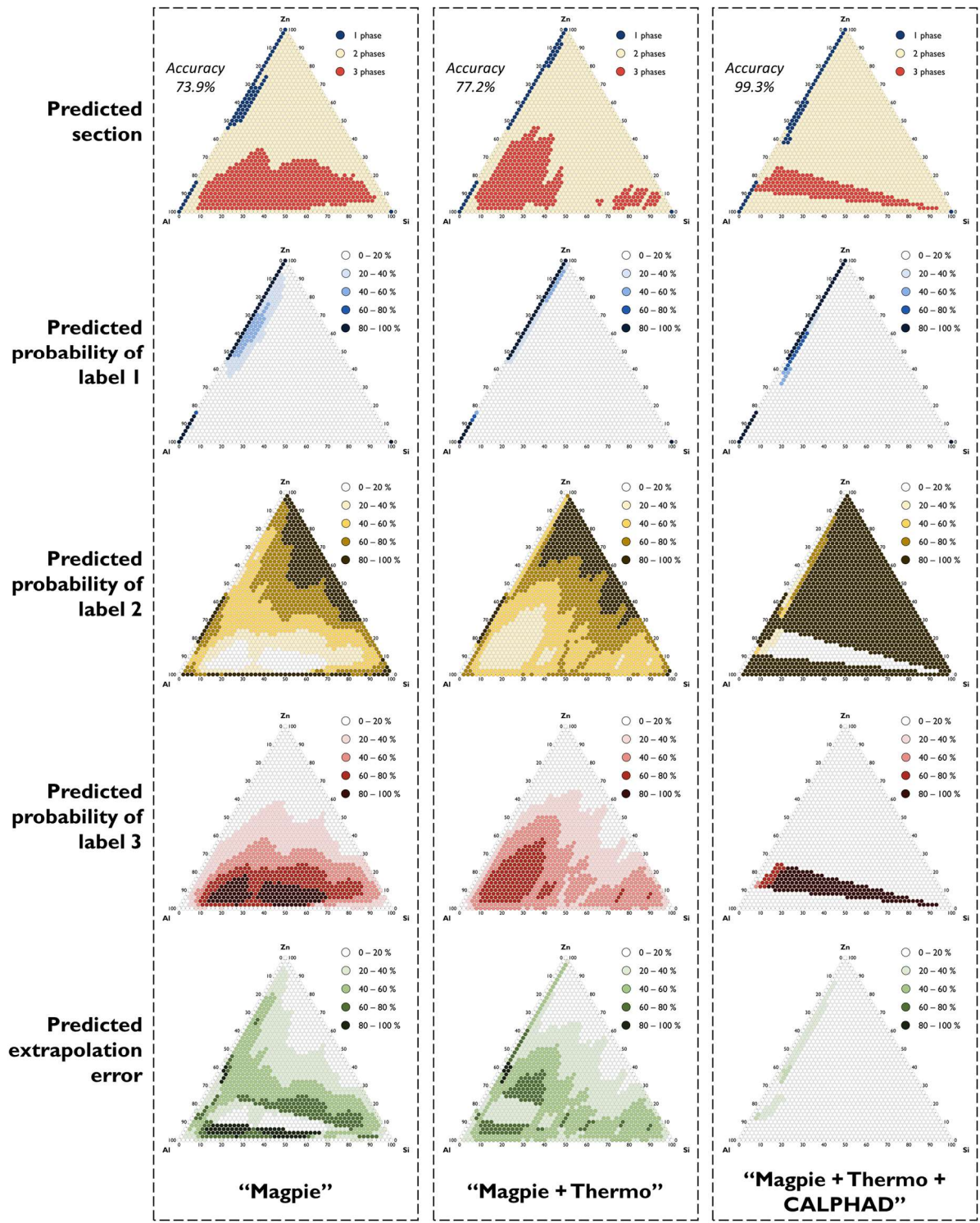

**Fig. S9.** Prediction of the isothermal section of the Al-Si-Zn system at 800 K.

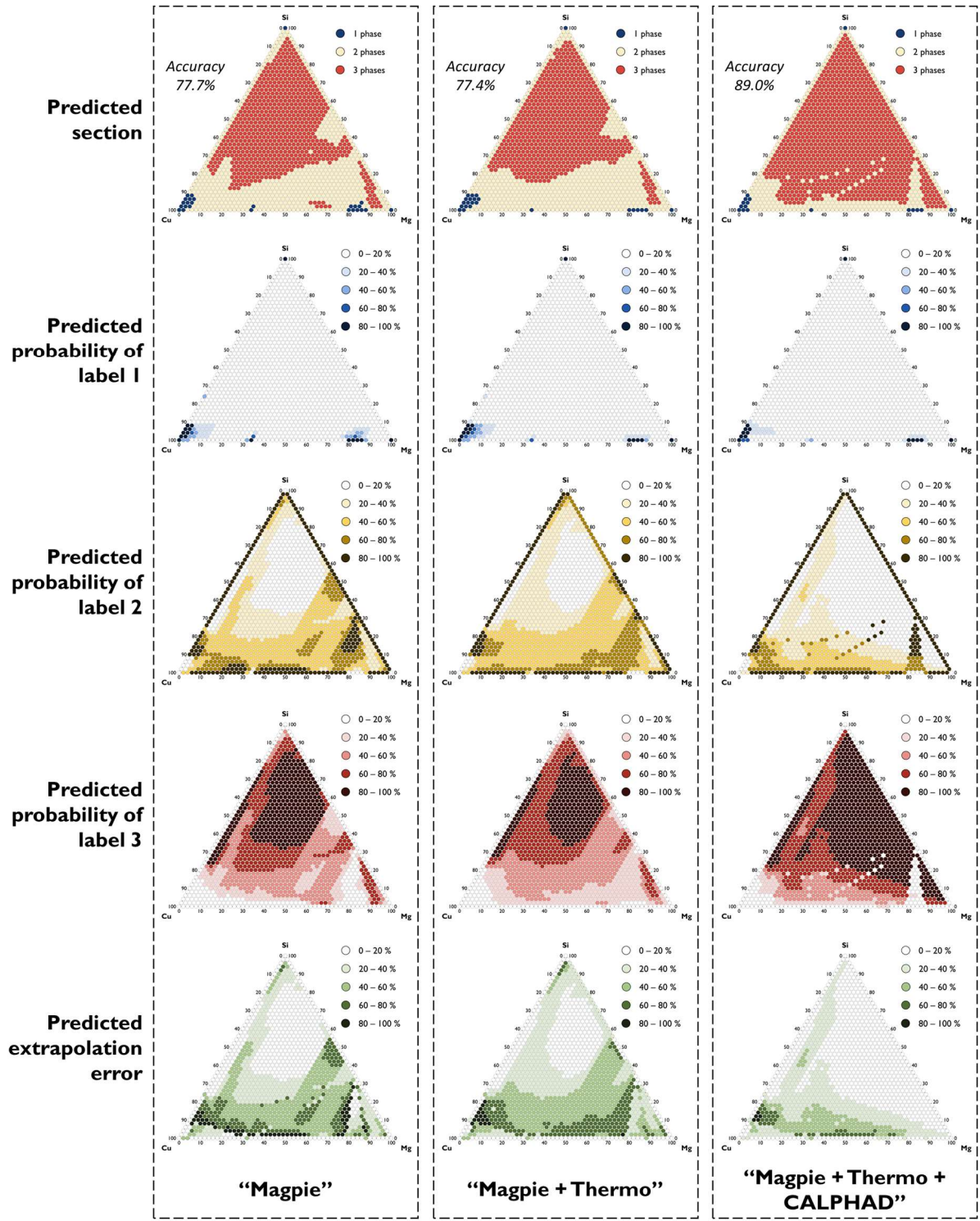

**Fig. S10.** Prediction of the isothermal section of the Cu-Mg-Si system at 800 K.

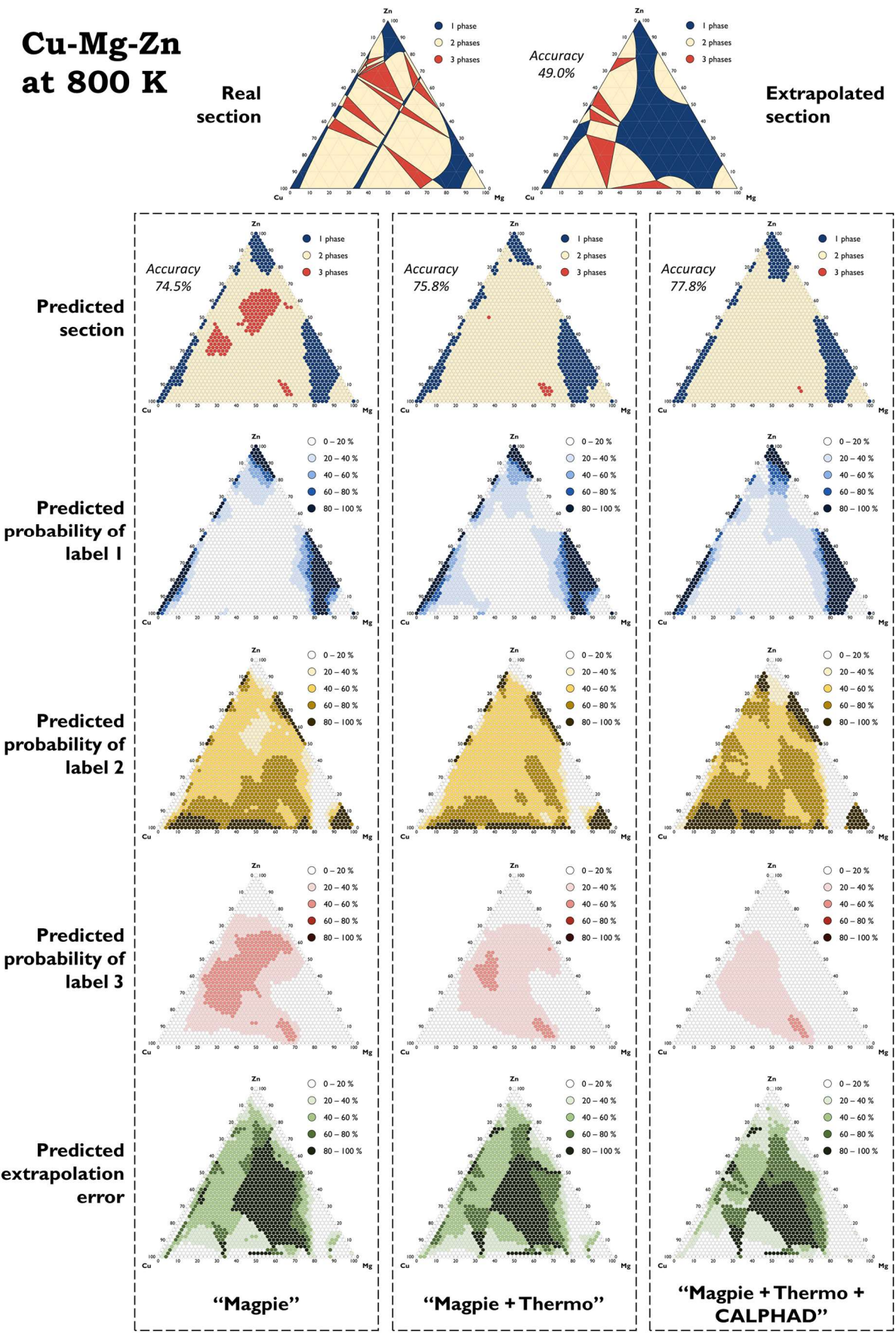

**Fig. S11.** Prediction of the isothermal section of the Cu-Mg-Zn system at 800 K.

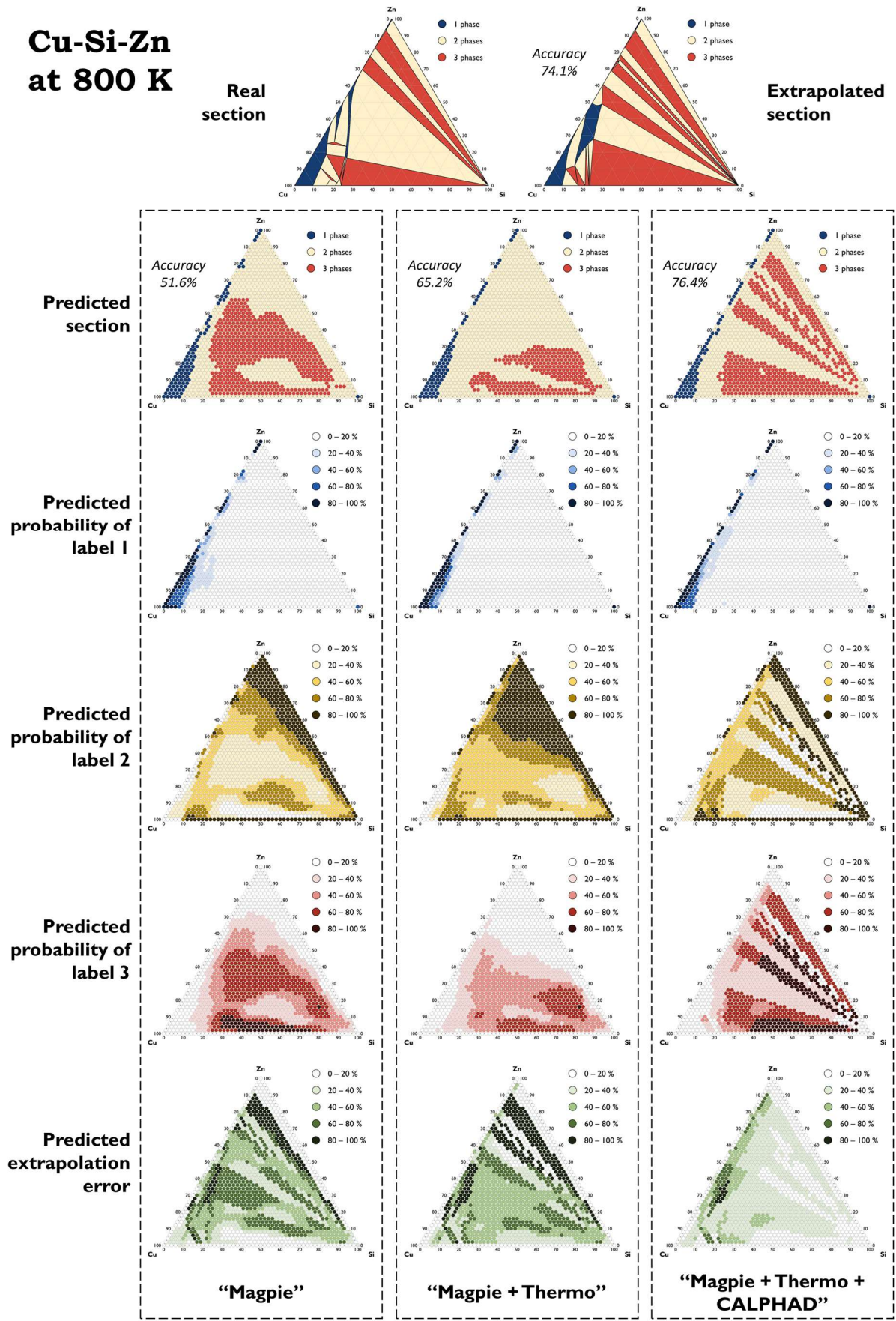

**Fig. S12.** Prediction of the isothermal section of the Cu-Si-Zn system at 800 K.

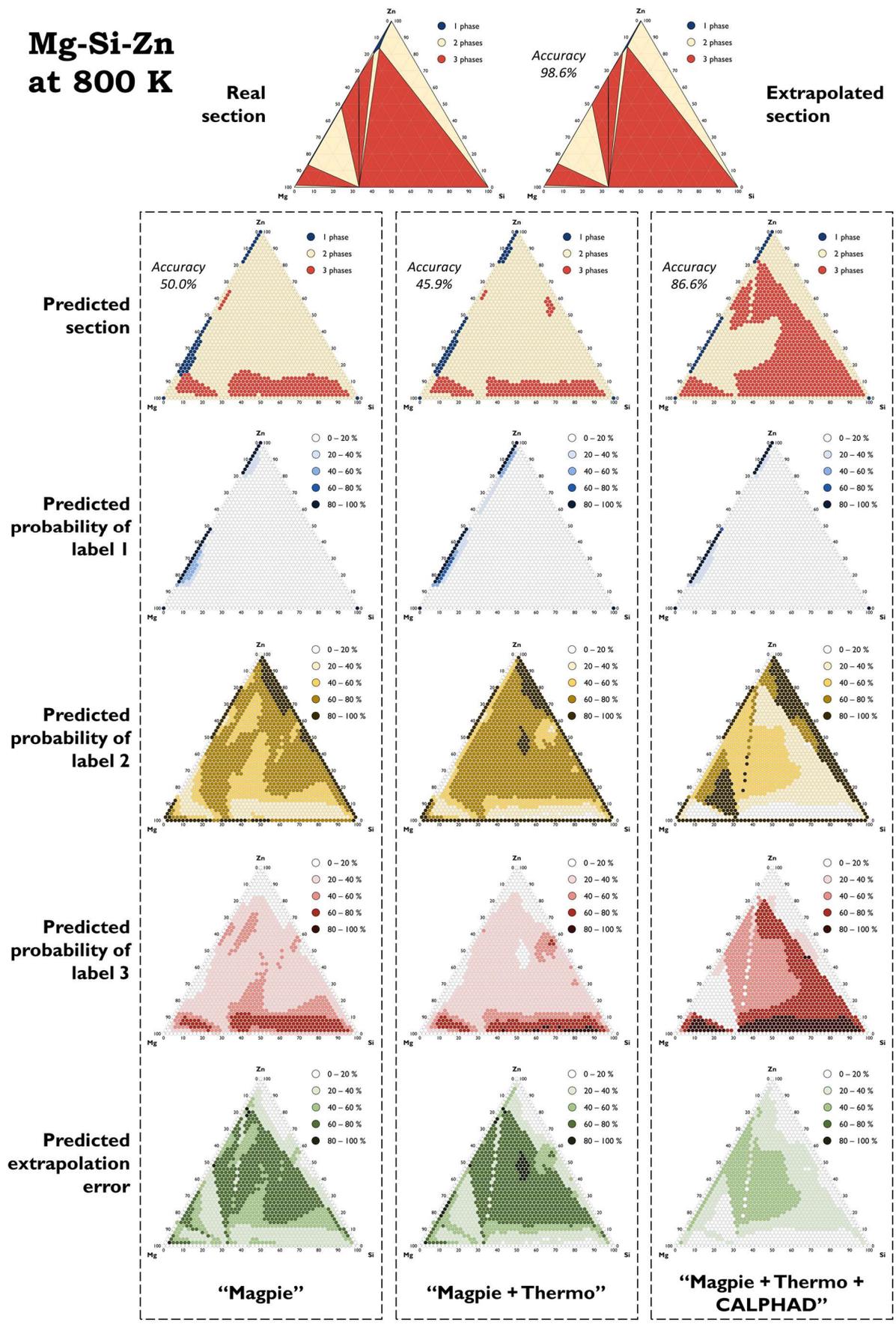

**Fig. S13.** Prediction of the isothermal section of the Mg-Si-Zn system at 800 K.